\documentclass[aps,prd,twocolumn,superscriptaddress,nofootinbib]{revtex4-1}

\pdfoutput=1

\usepackage{graphicx}
\usepackage{amssymb,amsmath,latexsym}
\usepackage{xcolor}
\usepackage[colorlinks=true,linktocpage=true,linkcolor=blue,citecolor=orange,urlcolor=blue]{hyperref}
\usepackage{subfigure}

\usepackage{soul}
\begin{document}
\title{Speed of sound peak in isospin QCD: A natural prediction of the medium separation scheme}
\author{Bruno S. Lopes} \email{bruno.lopes@acad.ufsm.br}
\affiliation{Departamento de F\'{i}sica, Universidade Federal de Santa Maria, 97105-900 Santa Maria, RS, Brazil}
\author{Dyana C. Duarte} \email{dyana.duarte@ufsm.br}
\affiliation{Departamento de F\'{i}sica, Universidade Federal de Santa Maria, 97105-900 Santa Maria, RS, Brazil}
\author{Ricardo L. S. Farias} \email{ricardo.farias@ufsm.br}
\affiliation{Departamento de F\'{i}sica, Universidade Federal de Santa Maria, 97105-900 Santa Maria, RS, Brazil}
\author{Rudnei O. Ramos} \email{rudnei@uerj.br}
\affiliation{Departamento de F\'{i}sica Te\'{o}rica, Universidade do Estado do Rio de Janeiro, 20550-013 Rio de Janeiro, RJ, Brazil}
%
%
\begin{abstract}

We present predictions for the zero-temperature equation of state at finite isospin density using the Nambu-Jona-Lasinio (NJL) 
model within the medium separation scheme (MSS)---a scheme that explicitly disentangles medium effects from the ultraviolet divergent vacuum terms. 
Recent lattice QCD results reveal a nonmonotonic speed of sound ($c_s^2$) as a function of isospin chemical potential ($\mu_I$), 
exhibiting explicit violation of the conformal bound $c_s^2 = 1/3$.
These findings have attracted significant theoretical interest, as established models---including the NJL model---failed to 
anticipate this behavior prior to lattice simulations. Conventional NJL implementations yield unphysical artifacts, 
often attributed to regularization scale sensitivity stemming from nonrenormalizability. However, in this work, we demonstrate 
that the standard NJL framework combined with MSS quantitatively reproduces state-of-the-art lattice data for isospin QCD.

\end{abstract}
\maketitle
%
%
\noindent {\it Introduction.} Over the years, significant efforts have been devoted to studying the phase structure of strongly interacting matter, as described by quantum chromodynamics (QCD). 
The current understanding of the QCD phase diagram at finite temperature ($T$) and baryon density ($n_B$) reflects a rich interplay between experimental and theoretical approaches. 
At low baryon density and increasing temperature, 
heavy-ion collisions and lattice QCD (LQCD) simulations indicate a smooth crossover from hadronic matter to 
the quark-gluon plasma~\cite{Gyulassy:2004zy,Borsanyi:2020fev}.

Exploration of the $T-n_B$ plane has been pursued through the beam energy scan (BES) program at the Relativistic Heavy-Ion Collider (RHIC)~\cite{Chen:2024aom}, 
with further investigations anticipated at upcoming facilities such as the Facility for Antiproton and Ion Research (FAIR)~\cite{CBM:2016kpk} and the Nuclotron-based Ion Collider fAcility (NICA)~\cite{MPD:2022qhn}. 
A key focus of these experiments is the search for the conjectured critical point, beyond which the phase transition is expected to become first-order.
Along with results from the detection of gravitational waves at LIGO/Virgo~\cite{LIGOScientific:2017vwq,LIGOScientific:2018cki} and observations by NICER~\cite{Miller:2019cac,Riley:2019yda,Miller:2021qha,Riley:2021pdl}, these perspectives have increased interest in studies of the high-density region of the phase diagram, particularly applied to the context of neutron stars. However, theoretical predictions of LQCD in this regime are hampered by the sign problem~\cite{deForcrand:2009zkb}, justifying the need for alternative approaches, such as the use of effective models.

The phase diagram can also be studied with additional dimensions, such as magnetic and electric fields~\cite{Adhikari:2024bfa,Endrodi:2024cqn,Klevansky:1989vi}, rotation effects~\cite{Jiang:2016woz}, and the isospin chemical potential, $\mu_I$. The latter is also relevant in neutron stars, and unlike the case at finite baryon density, LQCD simulations can be performed in the finite $T-\mu_I$ regime~\cite{Son:2000xc}, thus providing an important benchmark for effective models. 
These simulations revealed the formation of a pion superfluid as $\mu_I$ exceeds the pion mass, $m_\pi$, consistent with the model predictions 
available~\cite{Kogut:2002zg}. More recently, lattice results from different collaborations~\cite{Brandt:2022hwy,Abbott:2023coj,Abbott:2024vhj} 
have observed the formation of a peak structure in the speed of sound squared ($c_s^2$) with increasing isospin density, 
which can exceed the conformal limit of $1/3$. At large values of $\mu_I$, where perturbative QCD (pQCD) techniques are applicable, $c_s^2$
has been shown to approach the conformal limit from above in the presence of a superfluid gap~\cite{Fukushima:2024gmp}. 
As the peak existence was not previously predicted in model calculations, the subject has notoriously garnered significant interest in the literature. 
Since then, many approaches have been adopted  to capture this behavior in the context of effective models. {}For example, in Ref.~\cite{Chiba:2023ftg}, 
a quark-meson model (QMM) is considered, and the peak is linked to a crossover from the Bose-Einstein condensed phase to the Bardeen-Cooper-Schrieffer regime. In Ref.~\cite{Carlomagno:2024xmi}, a modified Nambu--Jona-Lasinio (NJL) model with nonlocal four-point interactions is investigated. The work in Ref.~\cite{Brandt:2025tkg} considered the two-flavor QMM within a renormalization group (RG) invariant approach. In a recent paper~\cite{Ivanytskyi:2025cnn}, it was suggested that a quarkyonic picture, including vector and isovector repulsion among the quark-antiquark-pion degrees of freedom, could be responsible for the peak behavior as the inclusion of nonlocal quark repulsion and a quark-based pion mass scaling ensures consistency with perturbative QCD, unlike approaches that treat pions as fundamental degrees of freedom. Additional studies include the investigations in Refs.~\cite{Kojo:2024sca,Chiba:2024cny,Andersen:2025ezj} and a holographic QCD approach~\cite{Kovensky:2024oqx}. In Ref.~\cite{Ayala:2023mms}, the effect of $\mu_I$-dependent couplings within the NJL model and two-flavor QMM was considered, while Ref.~\cite{Ayala:2024sqm} deals with the inclusion of one-loop meson contributions to the potential of the latter.

The peak in $c_s^2$ lying above the conformal limit seems to be a general feature of dense QCD, as suggested by finite baryon density phenomenology (astrophysical observations)~\cite{Bedaque:2014sqa,Fukushima:2025ujk}, two-color QCD studies~\cite{Iida:2024irv}, and also effects of the color superconducting phase~\cite{Gholami:2024ety}. In the case of finite isospin density, it was first reported from LQCD simulations in Ref.~\cite{Brandt:2022hwy}, with the peak in $c_s^2$ located at $\mu_I/m_\pi \approx 1.6$, with a maximum value $c_s^2 \approx 0.56$. This qualitative result was once again obtained from LQCD calculations in Refs.~\cite{Abbott:2023coj,Abbott:2024vhj}, although the location of the peak was shifted to higher densities at $\mu_I/m_\pi \approx 2.5$, while maintaining a maximum of similar height as the previous case. Within the same study, a unified model description combined the results of chiral effective theory ($\chi$EFT)~\cite{Adhikari:2019zaj}, continuum-extrapolated LQCD and pQCD~\cite{Kurkela:2009gj,Fujimoto:2023mvc} through a Gaussian process (GP). As such, we have an ideal scenario to assess the predictions of the isospin-asymmetric NJL model. 

In this paper, we show that the existence of the speed of sound peak can be simply explained in the context of the NJL effective model
when making use of the medium separation scheme (MSS). The MSS provides a way to fully disentangle medium contributions from divergent vacuum contributions 
such that only the latter are regularized. This prescription has proven reliable and often necessary for the correct description of various phenomenologies~\cite{Farias:2005cr,Farias:2016let,Duarte:2018kfd,Das:2019crc,Lopes:2021tro,Azeredo:2024sqc}.
We argue that the existence of the speed of sound peak is a natural prediction of the MSS, which in addition shows excellent agreement with LQCD data from Refs.~\cite{Abbott:2023coj,Abbott:2024vhj} and recovers the conformal limit at high densities, as opposed to the more standard way of regularizing
ultraviolet (UV) terms in NJL type of models\footnote{We call the usual form of regularizing UV divergent terms in the literature of traditional regularization scheme (TRS).}. 

The remainder of this paper is structured as follows. We first briefly recapitulate the formalism of the NJL model with an isospin imbalance. The results obtained from the TRS and MSS procedures are presented and discussed in sequence. Finally, we synthesize our findings and emphasize the importance of MSS in NJL-type models to properly describe medium-dependent physical quantities.

%
%
\bigskip \noindent {\it NJL Model within the Medium Separation Scheme.} The NJL model with an isospin asymmetry has been extensively discussed in the literature~\cite{Toublan:2003tt,Frank:2003ve,Barducci:2004tt,He:2005sp,He:2005nk,He:2006tn,Ebert:2005wr,Ebert:2005cs,Andersen:2007qv,Sun:2007fc,Abuki:2008wm,Mu:2010zz,Xia:2013caa,Ebert:2016hkd,Khunjua:2017khh,Khunjua:2018sro,Khunjua:2018jmn,Khunjua:2019lbv,Khunjua:2019ini,Avancini:2019ego,Lu:2019diy,Khunjua:2020xws,Lopes:2021tro,Khunjua:2021oxf,Liu:2021gsi,Ayala:2023mms,Carlomagno:2024xmi,Liu:2023uxm}. 
{}For further details on the medium separation scheme, see Refs.~\cite{Duarte:2018kfd,Lopes:2021tro}, which have considered  both a two-flavor and three-flavor analysis.

The partition function at finite temperature and chemical potential for a fermionic field theory model, with Lagrangian density
$\mathcal{L}$, is given by
\begin{align}
Z = \int [d\bar{\psi}][d\psi] \exp \left[ - \int_0^\beta \!\! d\tau \! \int \! d^3 x \left(\mathcal{L} + \hat{\mu}\bar{\psi}\gamma^0 \psi  \right) \right] \, ,
\end{align}
where $\psi$ are the quark spinors, $\hat{\mu}$ is the quark chemical potential matrix, and $\beta$ is the inverse temperature. In the case of the two-flavor model, the NJL Lagrangian density is given by
\begin{align}
\mathcal{L} = \bar{\psi} (i\gamma^\mu \partial_\mu - m_l)\psi + G\left[(\bar{\psi}\psi)^2 + (\bar{\psi}i\gamma^5\vec{\tau}\psi)^2\right] \, ,
\end{align}
in which $\psi=(u,d)^T$ in flavor space, $m_l$ is the current light quark mass, $G$ is the scalar/pseudoscalar interaction coupling and $\vec{\tau} = (\tau_1,\tau_2,\tau_3)$ are the Pauli matrices. {}For the three-flavor model, the group structure is modified and the Lagrangian density reads
\begin{align}
	\mathcal{L} &= \bar{\psi} \left(i\gamma^\mu\partial_\mu - \hat{m}\right)\psi + G \sum_{a=0}^{8} \left[ (\bar{\psi}\lambda_a\psi)^2 + (\bar{\psi}i\gamma_5\lambda_a\psi)^2  \right] \nonumber\\
	&- K \left\{\textrm{det}_f\left[\bar{\psi}\left(1 + \gamma_5\right)\psi\right] + \textrm{det}_f\left[\bar{\psi}\left(1 - \gamma_5\right)\psi\right]\right\} \, ,
\end{align}
where now $\psi = (u,d,s)^T$ includes the strange quark, $\hat{m} = {\rm diag}(m_u,m_d,m_s)$, $\lambda_a$ are the Gell-Mann matrices and $K$ is the coupling of the 't Hooft six-point interaction. We consider $m_u = m_d = m_l$ in the isospin symmetric approximation, at zero baryon and strangeness chemical potentials. The isospin chemical potential is introduced such that $\mu_u = \mu_I/2$ and $\mu_d = - \mu_I/2$. All results are obtained in the mean-field approximation, introducing the quark condensates $\left\langle \bar{u}u \right\rangle$, $\left\langle \bar{d}d \right\rangle$, and $\left\langle \bar{s}s \right\rangle$, which account for chiral symmetry breaking. At $\mu_I > m_\pi$ a pion condensate, $\Delta$, which is proportional to $\left\langle \bar{u}i\gamma^5 d \right\rangle$, is formed together with its conjugate, and they may differ by an arbitrary phase due to the spontaneous breaking of isospin symmetry~\cite{Xia:2013caa}.

The thermodynamic potential, obtained from the partition function as $\Omega = (\beta V)^{-1} \ln Z$, contains momentum integrals that are UV divergent and must be regularized. Given the nonrenormalizability of the NJL model, we introduce a cutoff $\Lambda$ that serves as an energy scale. The most common procedure in the literature (TRS), consists of simply introducing the UV cutoff $\Lambda$ in the divergent momentum integral. However, it is important to observe that the dispersion relations of the light quarks have the form
\begin{align}
E_k^{\pm} = \sqrt{\left(E_k \pm \frac{\mu_I}{2}\right)^2 + \Delta^2} ,
\label{Ek}
\end{align}
i.e., they are explicit functions of the isospin chemical potential. In Eq.~(\ref{Ek}), $E_k = \sqrt{k^2 + M_l^2}$ and $M_l$ is the effective quark mass. In TRS, these terms are integrated up to $k = \Lambda$, and important medium effects are lost. In the next section, it will be evident that this approach introduces artifacts to the model, and the results cannot be trusted, especially in the high density regime~\cite{Farias:2005cr,Farias:2016let,Duarte:2018kfd,Das:2019crc,Lopes:2021tro,Azeredo:2024sqc}. The MSS consists of manipulating the expressions for the gap equations~\footnote{The gap equations are obtained by extremizing the thermodynamic potential, $\frac{\partial\Omega}{\partial M} = \frac{\partial\Omega}{\partial \Delta} = 0$. All expressions for gap equations and thermodynamic potential at finite $\mu_I$ for $SU(2)$ and $SU(3)$ cases are shown in details in Refs.~\cite{Lopes:2021tro,Avancini:2019ego} for both TRS and MSS cases.}  and thermodynamic potential by adding and subtracting convenient terms, and then rearranging the expressions separating divergences in terms of vacuum quantities, namely the vacuum effective quark mass $M_0 = M(T = \mu_I = 0)$. Such divergent integrals in the scale $M_0$ can then be regularized with the momentum cutoff, while the remaining medium terms are finite and can be integrated without restrictions.

As we will see, regardless of the form of regularization employed, the MSS is crucial for a proper description of the thermodynamic quantities of interest, such as the pressure $P$, the isospin density $n_I$, the energy density $\epsilon$, and the speed of sound $c_s^2$, which are defined, respectively, as
\begin{align}
P &= - \left[\Omega(\mu_I) - \Omega(\mu_I = 0)\right] \, , \nonumber \\
n_I &= \frac{\partial P}{\partial \mu_I} \, , \nonumber \\
\epsilon &= - P + \mu_I n_I \, , \nonumber\\
c_s^2 &= \frac{\partial P}{\partial \epsilon} \, .
\end{align}
In the next section, we present the numerical results for each of these quantities.
It should be emphasized that an improper regularization of medium effects results in an incorrect behavior of thermodynamic quantities and may lead to nonphysical outcomes. A detailed discussion of causality violation and related issues arising from such improper treatments can be found in Ref.~\cite{Pasqualotto:2023hho}.


\bigskip \noindent {\it Numerical Results.} To compare our model predictions with the results from LQCD, in the following, we consider two representative sets of parameters: set I corresponds to a pion mass $m_\pi = 139.57$ MeV, used in the model of Ref.~\cite{Abbott:2024vhj}, while set II corresponds to $m_\pi = 169$ MeV to more closely represent the LQCD ensembles in Ref.~\cite{Abbott:2023coj}. The $SU(2)$ model takes as additional inputs the pion decay constant $f_\pi$ and the quark condensate $\left\langle\bar{u}u\right\rangle = \left\langle\bar{d}d\right\rangle$. The $SU(3)$ parametrization considers in addition to $f_\pi$ and $m_\pi$ also the kaon mass $m_K$ and the eta-meson mass $m_\eta$, while fixing the value of the current quark mass $m_l$. The parameters obtained with the sharp cutoff method are shown in Tables~\ref{tab:parameters_su2} and \ref{tab:parameters_su3},
for the $SU(2)$ and $SU(3)$ cases respectively. 
\begin{table}[h]
	\caption{Parameter sets for the $SU(2)$ NJL model.}
	\begin{center}
		\begin{tabular}{c c c}
			\hline Set & Input parameters  & Output parameters \\ \hline 
			& $f_\pi = 92.4$ MeV           & $\Lambda = 669.5$ MeV \\ 
		I	& $m_\pi = 139.57$ MeV         & $G = 2.029/\Lambda^2$ \\
			& $\left\langle\bar{u}u\right\rangle^{1/3} = - 250$ MeV  & $m_l = 5.226$ MeV  \\ \hline
			& $f_\pi = 92.4$ MeV           & $\Lambda = 674.6$ MeV\\  
		II	& $m_\pi = 169$ MeV            & $G =1.991/\Lambda^2$ \\
			& $\left\langle\bar{u}u\right\rangle^{1/3} = - 250$ MeV  & $m_l = 7.592$ MeV \\ \hline
		\end{tabular}
	\end{center}
	\label{tab:parameters_su2}
\end{table}
\begin{table}[h]
\caption{Parameter sets for the $SU(3)$ NJL model.}
\begin{center}
	\begin{tabular}{c c c}
	\hline Set & Input parameters & Output parameters \\ \hline 
			& $f_\pi = 92.4$ MeV           & $\Lambda = 643.9$ MeV \\ 
			& $m_\pi = 139.57$ MeV         & $G = 1.582/\Lambda^2$ \\ 
		I   & $m_K = 497.7$ MeV                & $K = 14.78/\Lambda^5$ \\ 
			& $m_\eta = 960.8$ MeV             & $m_s = 136.0$ MeV \\ 
			& $m_l= 5.5$ MeV               &  \\ \hline
			& $f_\pi = 92.4$ MeV           & $\Lambda = 663.0$ MeV\\  
			& $m_\pi = 169$ MeV            & $G =1.491/\Lambda^2$ \\ 
		II  & $m_K = 497.7$ MeV            & $K = 15.86/\Lambda^5$ \\ 
			& $m_{\eta^\prime}=960.8$ MeV  & $m_s= 131.3$ MeV \\ 
			& $m_l=7.8$ MeV                &  \\ \hline
	\end{tabular}
\end{center}
\label{tab:parameters_su3}
\end{table}
%
%
\begin{figure}[b]
\centering
\includegraphics[width=7.cm]{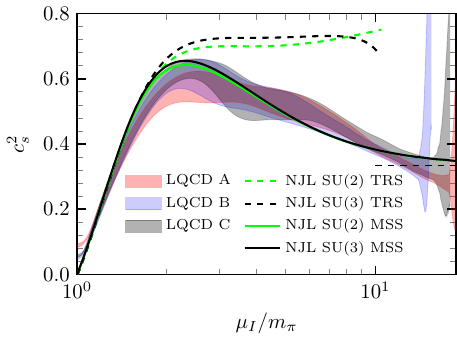}
\caption{Speed of sound $c_s^2$ as a function of the scaled isospin chemical potential $\mu_I/m_\pi$, comparing the LQCD ensembles from Refs.~\cite{Abbott:2023coj,Abbott:2024vhj} to NJL model results.}
\label{fig:cs2_ensembles}
\end{figure}
%
%
\begin{figure}[t]
\centering
\includegraphics[width=7.cm]{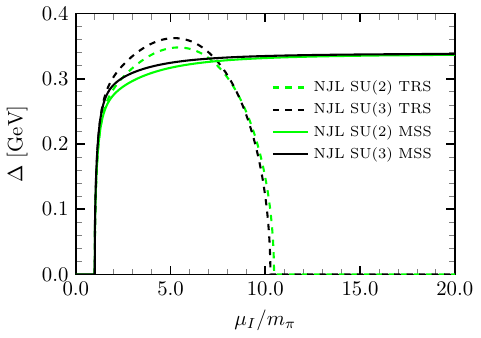}
\caption{Pion condensate $\Delta$ as a function of the scaled isospin chemical potential $\mu_I/m_\pi$, within both NJL regularization prescriptions.}
\label{fig:pion_condensate}
\end{figure}
%

In {}Fig.~\ref{fig:cs2_ensembles}, we show the behavior of the speed of sound as a function of the isospin chemical potential for both regularization prescriptions in the NJL model. Our results are compared with those obtained from the LQCD ensembles of Refs.~\cite{Abbott:2023coj,Abbott:2024vhj}. 
To more closely match the value of $m_\pi$ used in the lattice, we have adopted the parameter set II in this case. Evidently, the MSS results 
show excellent agreement with the LQCD data, lying almost entirely within the error bands, in addition to describing a peak of similar height and even converging to the conformal limit at large $\mu_I$. The TRS description, on the other hand, is not able to describe the overall behavior because a finite value cannot be assigned to the speed of sound at large isospin chemical potentials. This behavior can be related to the pion condensate, which vanishes at a high $\mu_I$ in the TRS, while remaining nonvanishing within the MSS prescription, as shown in {}Fig.~\ref{fig:pion_condensate}. The persistence of the pion condensate up to high values of $\mu_I$ in the MSS shown in {}Fig.~\ref{fig:pion_condensate} might be seen as an extrapolation, where we are going beyond the validity of the effective model. However, even at moderate values of $\mu_I$ we can clearly see evident differences between TRS and MSS. As we show next, these subtle differences between the schemes at moderate $\mu_I$ are enough to capture the correct behavior for $c_s^2$ in the MSS case. Nonetheless, Ref.~\cite{Abbott:2024vhj} estimates the pion condensate through a subtraction of the pQCD pressure without a gap from the LQCD pressure, finding that it still contributes significantly in the $\mu_I \in [1500,3000]$ MeV region.
%
%
\begin{figure}[b]
\centering
\includegraphics[width=7.cm]{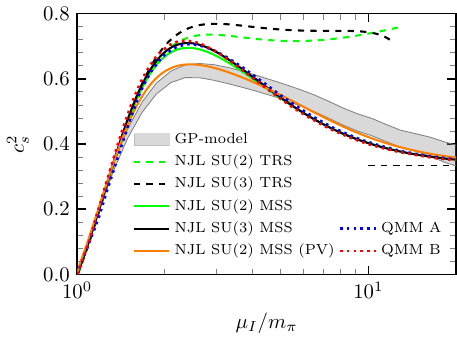}
\caption{Speed of sound $c_s^2$ as a function of the scaled isospin chemical potential $\mu_I/m_\pi$, comparing the GP-model from Ref.~\cite{Abbott:2024vhj} and the QMM from Ref.~\cite{Andersen:2025ezj} (A) and from Ref.~\cite{Brandt:2025tkg} (B) to the NJL model results.}
\label{fig:cs2_gpmodel}
\end{figure}
%

However, it should be noted that the reference LQCD data per ensemble may be subjected to finite lattice spacing effects and uncontrolled systematics at the low and high $\mu_I$ limits. {}For this reason, Ref.~\cite{Abbott:2024vhj} has considered a Bayesian model mixing framework, combining the results from pQCD, LQCD and $\chi$EFT through a Gaussian process in what is called the GP-model. A comparison between the NJL description and the GP-model is presented 
in {}Fig.~\ref{fig:cs2_gpmodel}. Since the corresponding pion mass is $m_\pi = 139.57$ MeV in this case, we utilize the parameter set I. Once more, we see that only the MSS prescription can reproduce the qualitative behavior, although this time the speed of sound goes slightly above the reference data near the peak, and decreases slightly faster at intermediate isospin chemical potentials before converging to the conformal limit. Interestingly enough, the QMM results from Ref.~\cite{Andersen:2025ezj} (QMM A) and Ref.~\cite{Brandt:2025tkg} (QMM B) show the same overshoot and undershoot profile, and are in excellent agreement with the NJL results within the MSS. The fact that results obtained in a renormalizable model in the form of the QMM can be replicated by the MSS is another strong argument in favor of this regularization prescription. Also notable is the similarity between the $SU(2)$ and $SU(3)$ model descriptions, indicating that the strange quark plays a small role in this system. This can be anticipated, as the isospin imbalance has origin in the light quark sector. In regard to the TRS results, we once again cannot verify the existence of the peak or the convergence to the conformal limit. This is again a consequence of the vanishing of the pion condensate at high values of $\mu_I$, as seen in the previous case for the TRS. 

\par In order to verify the dependence of our findings on the regularization method applied to the vacuum integrals, in {}Fig.~\ref{fig:cs2_gpmodel} we have also considered the result when using the Pauli-Villars (PV) regularization method within the MSS. We considered only the SU(2) case, which is representative enough as argued in the discussion of the results. {}For PV, the divergences can be of quadratic or logarithmic order in momentum. Hence, two additional terms are needed in the MSS scheme, more specifically, divergent loop functions are expressed as
\begin{align}
f(k,M_0) \to f(k,M_0) + C_1 f(k,M_1) + C_2 f(k,M_2) \, ,
\end{align}
where $M_i^2 = M_0^2 + \alpha_i \Lambda_{\rm PV}^2$. We choose $C_1 = 1$, $C_2 = -2$, $\alpha_1 = 2$ and $\alpha_2 = 1$, as commonly adopted in the literature.
The input parameters shown in Table~\ref{tab:parameters_su2} remain the same for PV and the corresponding outputs are: $\Lambda_{\rm PV} = 863.3~{\rm MeV}$, $G=3.708~{\rm GeV}^{-2}$ and $m_l = 5.205~{\rm MeV}$.
In {}Fig.~\ref{fig:cs2_gpmodel} we observe a good description of the GP-model data also for the PV-MSS case, with a slightly smoother curve than in the MSS-sharp cutoff case. Both the qualitative peak structure and the convergence to the conformal limit remain in the PV-MSS as well, thus indicating that the $c_s^2$ peak is a robust feature of MSS, rather than an artifact of a particular regularization of the vacuum. {}Furthermore, in all cases within the MSS, the height of the peak respects the upper bound $c_s^2 = 0.781$ of nuclear matter given in Ref.~\cite{Hippert:2024hum}, which was obtained through the calculation of hydrodynamic transport coefficients from first principles.

%
\begin{center}
\begin{figure}[!bth]
\subfigure[]{\includegraphics[width=7.cm]{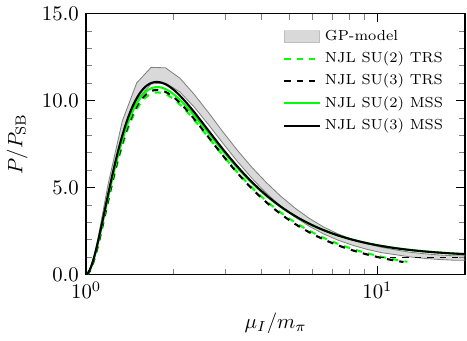}}
\subfigure[]{\includegraphics[width=7.cm]{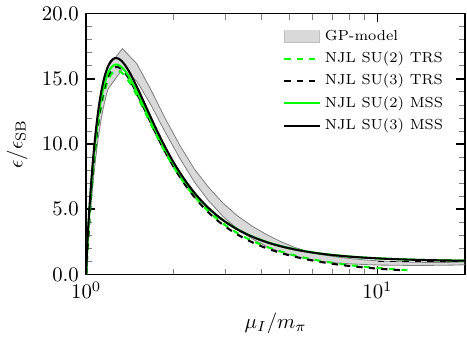}}
\caption{The pressure ratio $P/P_{\rm SB}$ (panel a) and the energy density ratio $\epsilon/\epsilon_{\rm SB}$ (panel b),
as a function of the scaled isospin chemical potential $\mu_I/m_\pi$.
The results are compared with the ones from the GP-model from Ref.~\cite{Abbott:2024vhj}.}
\label{fig:pres_gpmodel}
\end{figure}
\end{center}

{}For completeness, in the {}Fig.~\ref{fig:pres_gpmodel}(a) we present the pressure $P$ normalized by the Stefan-Boltzmann limit $P_{\rm SB} = \mu_I^4 / 32\pi^2$ as a function of the isospin chemical potential. Similarly, we show in the {}Fig.~\ref{fig:pres_gpmodel}(b) the energy density $\epsilon$ normalized by the ideal gas limit $\epsilon_{\rm SB} = 3\mu_I^4 / 32\pi^2$ as a function of $\mu_I$. The picture is very similar for both scaled quantities, with the existence of a peak at low isospin chemical potential and a convergence to the ideal limit at high $\mu_I$ within the MSS. Note that the location of the peak is slightly shifted compared to the one in the speed of sound. Overall, the NJL model description shows good agreement with the GP-model for the medium separation scheme, whereas the TRS fails to reach the Stefan-Boltzmann limits.


\bigskip \noindent {\it Conclusions.} In this paper, we investigated the importance of the regularization scheme for the description of isospin imbalanced matter in the context of the 
NJL model. Of particular interest is the existence of a peak in the speed of sound and recently observed by lattice QCD simulations~\cite{Brandt:2022hwy,Abbott:2023coj,Abbott:2024vhj}. We worked within the MSS, which disentangles medium contributions from the vacuum such that only the latter is regularized. The results were compared with those obtained by the usual TRS procedure, which is typically used in NJL-type models in the literature. 

Throughout the analysis, it was made clear that the speed of sound peak is a natural prediction of the MSS. One must keep in mind though that the NJL model has limited applicability at high energy scales as it is nonrenormalizable. Consequently, results with an isospin chemical potential greater than the regularization scale $\Lambda$ must be taken with great care. In fact, the TRS results are problematic at high $\mu_I$ and do not recover the conformal limit for the speed of sound. However, remarkably, 
despite the shortcomings expected from an effective model, the MSS results asymptotically approach $c_s^2 = 1/3$ as $\mu_I$ increases. This feature may be related to the behavior of the pion condensate, which is nonvanishing in this regime, in contrast to the prediction from the TRS prescription. Although concerned with superconducting systems, the renormalization group consistent regularization proposed in Ref.~\cite{Gholami:2024diy} finds a nonvanishing diquark condensate at large chemical potential, which is not the case with a conventional regularization. In this sense, the matching behaviors seem to signal that the MSS plays an important role in minimizing the dependence of physical quantities on the regularization scale. As the pion condensate is a central quantity in the analysis, a comparison with lattice results from Refs.~\cite{Brandt:2017oyy,Brandt:2018wkp} would be valuable. The aforementioned simulations, however, were performed at finite temperature, which requires a separate study from what we have done in the present paper \footnote{We note, however, that Ref.~\cite{Brandt:2018bwq} explored the case of temperatures
significantly below the relevant QCD scale, in which case the results presented in that reference can be well
approximated to $T=0$. Although the pion condensate is not explicitly presented there, relevant information can be extracted from their Fig.~3 and also using their Eq.~(A11).}.
An extension of this analysis that includes the pion condensate and also at finite temperature is being carried out and will be reported in future work.

In summary, the MSS has consistently demonstrated reliability across systems free from the sign problem, showing excellent agreement with lattice QCD where traditional cutoff-based schemes fail. Its successful applications extend to regimes inaccessible to lattice simulations, such as two-color-superconducting quark matter, where it yields results consistent with effective models. The present work advances this framework by showing that MSS naturally accounts for the peak in the speed of sound at finite isospin density, in agreement with recent findings from both lattice QCD and functional renormalization group approaches.
{}For this reason, we believe that a comparison with the RG-consistent method at finite isospin density would be interesting and could potentially help to further validate the applicability of the medium separation scheme also at the high density regime.

\bigskip \noindent {\it Acknowledgments.} This work was partially supported by Conselho Nacional de Desenvolvimento Cient\'ifico  e Tecno\-l\'o\-gico  (CNPq), Grants No. 312032/2023-4 (R.L.S.F.), No. 141270/2023-3 (B.S.L.) and No.  307286/2021-5 (R.O.R.); Fundação de Amparo à Pesquisa do Estado do Rio Grande do Sul (FAPERGS), Grants No. 24/2551-0001285-0 (R.L.S.F.), No. 23/2551-0000791-6, and No. 23/2551-0001591-9 (D.C.D); Funda\c{c}\~ao
Carlos Chagas Filho de Amparo \`a Pesquisa do Estado do Rio de Janeiro
(FAPERJ), Grant No. E-26/201.150/2021 (R.O.R.). The work is also part of the project Instituto
Nacional de Ciência e Tecnologia—Física Nuclear e
Aplicações (INCT—FNA), Grant No. 464898/2014-5
and supported by the Serrapilheira Institute (Grant
No. Serra-2211-42230).

\bigskip \noindent {\it Data Availability.} The data are not publicly available. The data are available from the authors upon reasonable request.
%

%

\begin{thebibliography}{99}

\bibitem{Gyulassy:2004zy}
M.~Gyulassy and L.~McLerran,
New forms of QCD matter discovered at RHIC,
\href{https://doi.org/10.1016/j.nuclphysa.2004.10.034}{Nucl. Phys. \textbf{A750}, 30 (2005)}.

\bibitem{Borsanyi:2020fev}
S.~Borsanyi, Z.~Fodor, J.~N.~Guenther, R.~Kara, S.~D.~Katz, P.~Parotto, A.~Pasztor, C.~Ratti, and K.~K.~Szabo,
QCD crossover at finite chemical potential from lattice simulations,
\href{https://doi.org/10.1103/PhysRevLett.125.052001}{Phys. Rev. Lett. \textbf{125}, 052001 (2020)}.

\bibitem{Chen:2024aom}
J.~Chen, J.~H.~Chen, X.~Dong, X.~He, X.~H.~He, H.~Huang, H.~Z.~Huang, F.~Liu, X.~Luo, X.~F.~Luo \textit{et al.},
Properties of the QCD matter: Review of selected results from the relativistic heavy ion collider beam energy scan (RHIC BES) program,
\href{https://doi.org/10.1007/s41365-024-01591-2}{Nucl. Sci. Tech. \textbf{35}, 214 (2024)}.

\bibitem{CBM:2016kpk}
T.~Ablyazimov \textit{et al.} (CBM Collaboration),
Challenges in QCD matter physics--The scientific programme of the compressed baryonic matter experiment at FAIR,
\href{https://doi.org/10.1140/epja/i2017-12248-y}{Eur. Phys. J. A \textbf{53}, 60 (2017)}.

\bibitem{MPD:2022qhn}
V.~Abgaryan \textit{et al.} (MPD Collaboration),
Status and initial physics performance studies of the MPD experiment at NICA,
\href{https://doi.org/10.1140/epja/s10050-022-00750-6}{Eur. Phys. J. A \textbf{58}, 140 (2022)}.

\bibitem{LIGOScientific:2017vwq}
B.~P.~Abbott \textit{et al.} (LIGO Scientific and Virgo Collaborations),
GW170817: Observation of gravitational waves from a binary neutron star inspiral,
\href{https://doi.org/10.1103/PhysRevLett.119.161101}{Phys. Rev. Lett. \textbf{119}, 161101 (2017)}.

\bibitem{LIGOScientific:2018cki}
B.~P.~Abbott \textit{et al.} (LIGO Scientific and Virgo Collaborations),
GW170817: Measurements of neutron star radii and equation of state,
\href{https://doi.org/10.1103/PhysRevLett.121.161101}{Phys. Rev. Lett. \textbf{121}, 161101 (2018)}.

\bibitem{Miller:2019cac}
M.~C.~Miller, F.~K.~Lamb, A.~J.~Dittmann, S.~Bogdanov, Z.~Arzoumanian, K.~C.~Gendreau, S.~Guillot, A.~K.~Harding, W.~C.~G.~Ho, J.~M.~Lattimer \textit{et al.},
PSR J0030+0451 mass and radius from $NICER$ data and implications for the properties of neutron star matter,
\href{https://doi.org/10.3847/2041-8213/ab50c5}{Astrophys. J. Lett. \textbf{887}, L24 (2019)}.

\bibitem{Riley:2019yda}
T.~E.~Riley, A.~L.~Watts, S.~Bogdanov, P.~S.~Ray, R.~M.~Ludlam, S.~Guillot, Z.~Arzoumanian, C.~L.~Baker, A.~V.~Bilous, D.~Chakrabarty \textit{et al.},
A NICER view of PSR J0030+0451: Millisecond pulsar parameter estimation,
\href{https://doi.org/10.3847/2041-8213/ab481c}{Astrophys. J. Lett. \textbf{887}, L21 (2019)}.

\bibitem{Miller:2021qha}
M.~C.~Miller, F.~K.~Lamb, A.~J.~Dittmann, S.~Bogdanov, Z.~Arzoumanian, K.~C.~Gendreau, S.~Guillot, W.~C.~G.~Ho, J.~M.~Lattimer, M.~Loewenstein \textit{et al.},
The radius of PSR J0740+6620 from NICER and XMM-Newton data,
\href{https://doi.org/10.3847/2041-8213/ac089b}{Astrophys. J. Lett. \textbf{918}, L28 (2021)}.

\bibitem{Riley:2021pdl}
T.~E.~Riley, A.~L.~Watts, P.~S.~Ray, S.~Bogdanov, S.~Guillot, S.~M.~Morsink, A.~V.~Bilous, Z.~Arzoumanian, D.~Choudhury, J.~S.~Deneva \textit{et al.},
A NICER view of the massive pulsar PSR J0740+6620 informed by radio timing and XMM-Newton spectroscopy,
\href{https://doi.org/10.3847/2041-8213/ac0a81}{Astrophys. J. Lett. \textbf{918}, L27 (2021)}.

\bibitem{deForcrand:2009zkb}
P.~de Forcrand,
Simulating QCD at finite density,
\href{https://doi.org/10.22323/1.091.0010}{Proc. Sci. LAT2009 \textbf{(2009)} 010}
[\href{https://arxiv.org/abs/1005.0539}{arXiv:1005.0539}].

\bibitem{Adhikari:2024bfa}
P.~Adhikari, M.~Ammon, S.~S.~Avancini, A.~Ayala, A.~Bandyopadhyay, D.~Blaschke, F.~L.~Braghin, P.~Buividovich, R.~P.~Cardoso, C.~Cartwright \textit{et al.},
Strongly interacting matter in extreme magnetic fields,
\href{https://doi.org/10.1016/j.ppnp.2025.104199}{Prog. Part. Nucl. Phys. \textbf{146}, 104199 (2026)}.

\bibitem{Endrodi:2024cqn}
G.~Endrodi,
QCD with background electromagnetic fields on the lattice: A review,
\href{https://doi.org/10.1016/j.ppnp.2024.104153}{Prog. Part. Nucl. Phys. \textbf{141}, 104153 (2025)}.

\bibitem{Klevansky:1989vi}
S.~P.~Klevansky and R.~H.~Lemmer,
Chiral symmetry restoration in the Nambu-Jona-Lasinio model with a constant electromagnetic field,
\href{https://doi.org/10.1103/PhysRevD.39.3478}{Phys. Rev. D \textbf{39}, 3478 (1989)}.

\bibitem{Jiang:2016woz}
Y.~Jiang, Z.~W.~Lin, and J.~Liao,
Rotating quark-gluon plasma in relativistic heavy ion collisions,
\href{https://doi.org/10.1103/PhysRevC.94.044910}{Phys. Rev. C \textbf{94}, 044910 (2016)};
\href{https://doi.org/10.1103/PhysRevC.95.049904}{\textbf{95}, 049904(E) (2017)}.

\bibitem{Son:2000xc}
D.~T.~Son and M.~A.~Stephanov,
QCD at finite isospin density,
\href{https://doi.org/10.1103/PhysRevLett.86.592}{Phys. Rev. Lett. \textbf{86}, 592 (2001)}.

\bibitem{Kogut:2002zg}
J.~B.~Kogut and D.~K.~Sinclair,
Lattice QCD at finite isospin density at zero and finite temperature,
\href{https://doi.org/10.1103/PhysRevD.66.034505}{Phys. Rev. D \textbf{66}, 034505 (2002)}.

\bibitem{Brandt:2022hwy}
B.~B.~Brandt, F.~Cuteri, and G.~Endrodi,
Equation of state and speed of sound of isospin-asymmetric QCD on the lattice,
\href{https://doi.org/10.1007/JHEP07(2023)055}{J. High Energy Phys. 07 (2023) 055}.

\bibitem{Abbott:2023coj}
R.~Abbott \textit{et al.} (NPLQCD Collaboration),
Lattice quantum chromodynamics at large isospin density,
\href{https://doi.org/10.1103/PhysRevD.108.114506}{Phys. Rev. D \textbf{108}, 114506 (2023)}.

\bibitem{Abbott:2024vhj}
R.~Abbott \textit{et al.} (NPLQCD Collaboration),
QCD constraints on isospin-dense matter and the nuclear equation of state,
\href{https://doi.org/10.1103/PhysRevLett.134.011903}{Phys. Rev. Lett. \textbf{134}, 011903 (2025)}.

\bibitem{Fukushima:2024gmp}
K.~Fukushima and S.~Minato,
Speed of sound and trace anomaly in a unified treatment of the two-color diquark superfluid, the pion-condensed high-isospin matter, and the 2SC quark matter,
\href{https://doi.org/10.1103/PhysRevD.111.094006}{Phys. Rev. D \textbf{111}, 094006 (2025)}.

\bibitem{Chiba:2023ftg}
R.~Chiba and T.~Kojo,
Sound velocity peak and conformality in isospin QCD,
\href{https://doi.org/10.1103/PhysRevD.109.076006}{Phys. Rev. D \textbf{109}, 076006 (2024)}.

\bibitem{Carlomagno:2024xmi}
J.~P.~Carlomagno, D.~Gomez Dumm, and N.~N.~Scoccola,
Cold isospin asymmetric baryonic rich matter in nonlocal NJL-like models,
\href{https://doi.org/10.1103/PhysRevD.109.094041}{Phys. Rev. D \textbf{109}, 094041 (2024)}.

\bibitem{Brandt:2025tkg}
B.~B.~Brandt, V.~Chelnokov, G.~Endrodi, G.~Marko, D.~Scheid, and L.~von Smekal,
Renormalization group invariant mean-field model for QCD at finite isospin density,
\href{https://doi.org/10.1103/fryz-f3vw}{Phys. Rev. D \textbf{112}, 054038 (2025)}.

\bibitem{Ivanytskyi:2025cnn}
O.~Ivanytskyi,
Quarkyonic picture of isospin QCD,
\href{https://doi.org/10.1103/831v-8mp4}{Phys. Rev. D \textbf{112}, 034001 (2025)}.

\bibitem{Kojo:2024sca}
T.~Kojo, D.~Suenaga, and R.~Chiba,
Isospin QCD as a laboratory for dense QCD,
\href{https://doi.org/10.3390/universe10070293}{Universe \textbf{10}, 293 (2024)}.

\bibitem{Chiba:2024cny}
R.~Chiba, T.~Kojo, and D.~Suenaga,
Thermal effects on sound velocity peak and conformality in isospin QCD,
\href{https://doi.org/10.1103/PhysRevD.110.054037}{Phys. Rev. D \textbf{110}, 054037 (2024)}.

\bibitem{Andersen:2025ezj}
J.~O.~Andersen and M.~P.~N{\o}dtvedt,
Pion condensation versus 2SC, speed of sound, and charge neutrality effects in the quark-meson diquark model,
\href{https://arxiv.org/abs/2502.10229}{arXiv:2502.10229}.

\bibitem{Kovensky:2024oqx}
N.~Kovensky and A.~Schmitt,
Thermal pion condensation: Holography meets lattice QCD,
\href{https://doi.org/10.1007/JHEP10(2024)133}{J. High Energy Phys. 10 (2024) 133}.

\bibitem{Ayala:2023mms}
A.~Ayala, B.~S.~Lopes, R.~L.~S.~Farias, and L.~C.~Parra,
Describing the speed of sound peak of isospin-asymmetric cold strongly interacting matter using effective models,
\href{https://doi.org/10.1140/epja/s10050-024-01469-2}{Eur. Phys. J. A \textbf{60}, 250 (2024)}.

\bibitem{Ayala:2024sqm}
A.~Ayala, B.~S.~Lopes, R.~L.~S.~Farias, and L.~C.~Parra,
On the origin of the peak of the sound velocity for isospin imbalanced strongly interacting matter,
\href{https://doi.org/10.1016/j.physletb.2025.139396}{Phys. Lett. B \textbf{864}, 139396 (2025)}.

\bibitem{Bedaque:2014sqa}
P.~Bedaque and A.~W.~Steiner,
Sound velocity bound and neutron stars,
\href{https://doi.org/10.1103/PhysRevLett.114.031103}{Phys. Rev. Lett. \textbf{114}, 031103 (2015)}.

\bibitem{Fukushima:2025ujk}
K.~Fukushima,
QCD phase diagram and astrophysical implications,
\href{https://doi.org/10.1016/j.jspc.2025.100066}{J. Subatomic Part. Cosmol. \textbf{3}, 100066 (2025)}.

\bibitem{Iida:2024irv}
K.~Iida, E.~Itou, K.~Murakami, and D.~Suenaga,
Lattice study on finite density QC$_{2}$D towards zero temperature,
\href{https://doi.org/10.1007/JHEP10(2024)022}{J. High Energy Phys. 10 (2024) 022}.

\bibitem{Gholami:2024ety}
H.~Gholami, I.~A.~Rather, M.~Hofmann, M.~Buballa, and J.~Schaffner-Bielich,
Astrophysical constraints on color-superconducting phases in compact stars within the RG-consistent NJL model,
\href{https://doi.org/10.1103/PhysRevD.111.103034}{Phys. Rev. D \textbf{111}, 103034 (2025)}.

\bibitem{Adhikari:2019zaj}
P.~Adhikari and J.~O.~Andersen,
QCD at finite isospin density: Chiral perturbation theory confronts lattice data,
\href{https://doi.org/10.1016/j.physletb.2020.135352}{Phys. Lett. B \textbf{804}, 135352 (2020)}.

\bibitem{Kurkela:2009gj}
A.~Kurkela, P.~Romatschke, and A.~Vuorinen,
Cold quark matter,
\href{https://doi.org/10.1103/PhysRevD.81.105021}{Phys. Rev. D \textbf{81}, 105021 (2010)}.

\bibitem{Fujimoto:2023mvc}
Y.~Fujimoto,
Enhanced contribution of the pairing gap to the QCD equation of state at large isospin chemical potential,
\href{https://doi.org/10.1103/PhysRevD.109.054035}{Phys. Rev. D \textbf{109}, 054035 (2024)}.

\bibitem{Farias:2005cr}
R.~L.~S.~Farias, G.~Dallabona, G.~Krein, and O.~A.~Battistel,
Cutoff-independent regularization of four-fermion interactions for color superconductivity,
\href{https://doi.org/10.1103/PhysRevC.73.018201}{Phys. Rev. C \textbf{73}, 018201 (2006)}.

\bibitem{Farias:2016let}
R.~L.~S.~Farias, D.~C.~Duarte, G.~Krein, and R.~O.~Ramos,
Thermodynamics of quark matter with a chiral imbalance,
\href{https://doi.org/10.1103/PhysRevD.94.074011}{Phys. Rev. D \textbf{94}, 074011 (2016)}.

\bibitem{Duarte:2018kfd}
D.~C.~Duarte, R.~L.~S.~Farias, and R.~O.~Ramos,
Regularization issues for a cold and dense quark matter model in $\beta-$equilibrium,
\href{https://doi.org/10.1103/PhysRevD.99.016005}{Phys. Rev. D \textbf{99}, 016005 (2019)}.

\bibitem{Das:2019crc}
A.~Das, D.~Kumar, and H.~Mishra,
Chiral susceptibility in the Nambu{\textendash}Jona-Lasinio model: A Wigner function approach,
\href{https://doi.org/10.1103/PhysRevD.100.094030}{Phys. Rev. D \textbf{100}, 094030 (2019)}.

\bibitem{Lopes:2021tro}
B.~S.~Lopes, S.~S.~Avancini, A.~Bandyopadhyay, D.~C.~Duarte, and R.~L.~S.~Farias,
Hot QCD at finite isospin density: Confronting the SU(3) Nambu{\textendash}Jona-Lasinio model with recent lattice data,
\href{https://doi.org/10.1103/PhysRevD.103.076023}{Phys. Rev. D \textbf{103}, 076023 (2021)}.

\bibitem{Azeredo:2024sqc}
F.~X.~Azeredo, D.~C.~Duarte, R.~L.~S.~Farias, G.~Krein, and R.~O.~Ramos,
Deconfinement and chiral phase transitions in quark matter with chiral imbalance,
\href{https://doi.org/10.1103/PhysRevD.110.076007}{Phys. Rev. D \textbf{110}, 076007 (2024)}.

\bibitem{Toublan:2003tt}
D.~Toublan and J.~B.~Kogut,
Isospin chemical potential and the QCD phase diagram at nonzero temperature and baryon chemical potential,
\href{https://doi.org/10.1016/S0370-2693(03)00701-9}{Phys. Lett. B \textbf{564}, 212 (2003)}.

\bibitem{Frank:2003ve}
M.~Frank, M.~Buballa, and M.~Oertel,
Flavor mixing effects on the QCD phase diagram at nonvanishing isospin chemical potential: One or two phase transitions?,
\href{https://doi.org/10.1016/S0370-2693(03)00607-5}{Phys. Lett. B \textbf{562}, 221 (2003)}.

\bibitem{Barducci:2004tt}
A.~Barducci, R.~Casalbuoni, G.~Pettini, and L.~Ravagli,
A calculation of the QCD phase diagram at finite temperature, and baryon and isospin chemical potentials,
\href{https://doi.org/10.1103/PhysRevD.69.096004}{Phys. Rev. D \textbf{69}, 096004 (2004)}.

\bibitem{He:2005sp}
L.~He and P.~Zhuang,
Phase structure of Nambu-Jona-Lasinio model at finite isospin density,
\href{https://doi.org/10.1016/j.physletb.2005.03.066}{Phys. Lett. B \textbf{615}, 93 (2005)}.

\bibitem{He:2005nk}
L.~y.~He, M.~Jin, and P.~f.~Zhuang,
Pion superfluidity and meson properties at finite isospin density,
\href{https://doi.org/10.1103/PhysRevD.71.116001}{Phys. Rev. D \textbf{71}, 116001 (2005)}.

\bibitem{He:2006tn}
L.~He, M.~Jin, and P.~Zhuang,
Pion condensation in baryonic matter: From Sarma phase to Larkin-Ovchinnikov-Fudde-Ferrell phase,
\href{https://doi.org/10.1103/PhysRevD.74.036005}{Phys. Rev. D \textbf{74}, 036005 (2006)}.

\bibitem{Ebert:2005wr}
D.~Ebert and K.~G.~Klimenko,
Pion condensation in electrically neutral cold matter with finite baryon density,
\href{https://doi.org/10.1140/epjc/s2006-02527-5}{Eur. Phys. J. C \textbf{46}, 771 (2006)}.

\bibitem{Ebert:2005cs}
D.~Ebert and K.~G.~Klimenko,
Gapless pion condensation in quark matter with finite baryon density,
\href{https://doi.org/10.1088/0954-3899/32/5/001}{J. Phys. G \textbf{32}, 599 (2006)}.

\bibitem{Andersen:2007qv}
J.~O.~Andersen and L.~Kyllingstad,
Pion condensation in a two-flavor NJL model: The role of charge neutrality,
\href{https://doi.org/10.1088/0954-3899/37/1/015003}{J. Phys. G \textbf{37}, 015003 (2009)}.

\bibitem{Sun:2007fc}
G.~f.~Sun, L.~He, and P.~Zhuang,
BEC-BCS crossover in the Nambu-Jona-Lasinio model of QCD,
\href{https://doi.org/10.1103/PhysRevD.75.096004}{Phys. Rev. D \textbf{75}, 096004 (2007)}.

\bibitem{Abuki:2008wm}
H.~Abuki, R.~Anglani, R.~Gatto, M.~Pellicoro, and M.~Ruggieri,
The fate of pion condensation in quark matter: From the chiral to the real world,
\href{https://doi.org/10.1103/PhysRevD.79.034032}{Phys. Rev. D \textbf{79}, 034032 (2009)}.

\bibitem{Mu:2010zz}
C.~f.~Mu, L.~y.~He, and Y.~x.~Liu,
Evaluating the phase diagram at finite isospin and baryon chemical potentials in the Nambu-Jona-Lasinio model,
\href{https://doi.org/10.1103/PhysRevD.82.056006}{Phys. Rev. D \textbf{82}, 056006 (2010)}.

\bibitem{Xia:2013caa}
T.~Xia, L.~He, and P.~Zhuang,
Three-flavor Nambu{\textendash}Jona-Lasinio model at finite isospin chemical potential,
\href{https://doi.org/10.1103/PhysRevD.88.056013}{Phys. Rev. D \textbf{88}, 056013 (2013)}.

\bibitem{Ebert:2016hkd}
D.~Ebert, T.~G.~Khunjua, and K.~G.~Klimenko,
Duality between chiral symmetry breaking and charged pion condensation at large $N_c$: Consideration of an NJL$_2$ model with baryon, isospin, and chiral isospin chemical potentials,
\href{https://doi.org/10.1103/PhysRevD.94.116016}{Phys. Rev. D \textbf{94}, no.11, 116016 (2016)}.

\bibitem{Khunjua:2017khh}
T.~G.~Khunjua, K.~G.~Klimenko, R.~N.~Zhokhov, and V.~C.~Zhukovsky,
Inhomogeneous charged pion condensation in chiral asymmetric dense quark matter in the framework of NJL$_2$ model,
\href{https://doi.org/10.1103/PhysRevD.95.105010}{Phys. Rev. D \textbf{95}, 105010 (2017)}.

\bibitem{Khunjua:2018sro}
T.~G.~Khunjua, K.~G.~Klimenko, and R.~N.~Zhokhov,
Dualities in dense quark matter with isospin, chiral, and chiral isospin imbalance in the framework of the large-N$_{c}$ limit of the NJL$_{4}$ model,
\href{https://doi.org/10.1103/PhysRevD.98.054030}{Phys. Rev. D \textbf{98}, 054030 (2018)}.

\bibitem{Khunjua:2018jmn}
T.~G.~Khunjua, K.~G.~Klimenko, and R.~N.~Zhokhov,
Chiral imbalanced hot and dense quark matter: NJL analysis at the physical point and comparison with lattice QCD,
\href{https://doi.org/10.1140/epjc/s10052-019-6654-2}{Eur. Phys. J. C \textbf{79}, 151 (2019)}.

\bibitem{Khunjua:2019lbv}
T.~G.~Khunjua, K.~G.~Klimenko, and R.~N.~Zhokhov,
Dualities and inhomogeneous phases in dense quark matter with chiral and isospin imbalances in the framework of effective model,
\href{https://doi.org/10.1007/JHEP06(2019)006}{J. High Energy Phys. 06 (2019) 006}.

\bibitem{Khunjua:2019ini}
T.~G.~Khunjua, K.~G.~Klimenko, and R.~N.~Zhokhov,
Charged pion condensation and duality in dense and hot chirally and isospin asymmetric quark matter in the framework of the NJL$_2$ model,
\href{https://doi.org/10.1103/PhysRevD.100.034009}{Phys. Rev. D \textbf{100}, 034009 (2019)}.

\bibitem{Avancini:2019ego}
S.~S.~Avancini, A.~Bandyopadhyay, D.~C.~Duarte, and R.~L.~S.~Farias,
Cold QCD at finite isospin density: Confronting effective models with recent lattice data,
\href{https://doi.org/10.1103/PhysRevD.100.116002}{Phys. Rev. D \textbf{100}, 116002 (2019)}.

\bibitem{Lu:2019diy}
Z.~Y.~Lu, C.~J.~Xia, and M.~Ruggieri,
Thermodynamics and susceptibilities of isospin imbalanced QCD matter,
\href{https://doi.org/10.1140/epjc/s10052-020-7614-6}{Eur. Phys. J. C \textbf{80}, 46 (2020)}.

\bibitem{Khunjua:2020xws}
T.~G.~Khunjua, K.~G.~Klimenko, and R.~N.~Zhokhov,
The dual properties of chiral and isospin asymmetric dense quark matter formed of two-color quarks,
\href{https://doi.org/10.1007/JHEP06(2020)148}{J. High Energy Phys. 06 (2020) 148}.

\bibitem{Khunjua:2021oxf}
T.~G.~Khunjua, K.~G.~Klimenko, and R.~N.~Zhokhov,
Influence of chiral chemical potential {\ensuremath{\mu}}5 on phase structure of the two-color quark matter,
\href{https://doi.org/10.1103/PhysRevD.106.045008}{Phys. Rev. D \textbf{106}, 045008 (2022)}.

\bibitem{Liu:2021gsi}
L.~M.~Liu, J.~Xu, and G.~X.~Peng,
Three-dimensional QCD phase diagram with a pion condensate in the NJL model,
\href{https://doi.org/10.1103/PhysRevD.104.076009}{Phys. Rev. D \textbf{104}, 076009 (2021)}.

\bibitem{Liu:2023uxm}
L.~M.~Liu, J.~Xu, and G.~X.~Peng,
Three-dimensional QCD phase diagram in the pNJL model,
\href{https://doi.org/10.11804/NuclPhysRev.40.2023025}{Nucl. Phys. Rev. \textbf{40}, 493 (2023)}.

\bibitem{Pasqualotto:2023hho}
A.~E.~B.~Pasqualotto, R.~L.~S.~Farias, W.~R.~Tavares, S.~S.~Avancini, and G.~Krein,
Causality violation and the speed of sound of hot and dense quark matter in the Nambu{\textendash}Jona-Lasinio model,
\href{https://doi.org/10.1103/PhysRevD.107.096017}{Phys. Rev. D \textbf{107}, 096017 (2023)}.


\bibitem{Hippert:2024hum}
M.~Hippert, J.~Noronha, and P.~Romatschke,
Upper bound on the speed of sound in nuclear matter from transport,
\href{https://doi.org/10.1016/j.physletb.2024.139184}{Phys. Lett. B \textbf{860}, 139184 (2025)}.

\bibitem{Gholami:2024diy}
H.~Gholami, M.~Hofmann, and M.~Buballa,
Renormalization-group consistent treatment of color superconductivity in the NJL model,
\href{https://doi.org/10.1103/PhysRevD.111.014006}{Phys. Rev. D \textbf{111}, 014006 (2025)}.

\bibitem{Brandt:2017oyy}
B.~B.~Brandt, G.~Endrodi, and S.~Schmalzbauer,
QCD phase diagram for nonzero isospin-asymmetry,
\href{https://doi.org/10.1103/PhysRevD.97.054514}{Phys. Rev. D \textbf{97}, 054514 (2018)}.

\bibitem{Brandt:2018wkp}
B.~B.~Brandt, G.~Endrodi, and S.~Schmalzbauer,
QCD at nonzero isospin asymmetry,
\href{https://doi.org/10.22323/1.336.0260}{Proc. Sci. Confinement2018 (\textbf{2018}) 260}
[\href{https://arxiv.org/abs/1811.06004}{arXiv:1811.06004}].

\bibitem{Brandt:2018bwq}
B.~B.~Brandt, G.~Endrodi, E.~S.~Fraga, M.~Hippert, J.~Schaffner-Bielich, and S.~Schmalzbauer,
New class of compact stars: Pion stars,
\href{https://doi.org/10.1103/PhysRevD.98.094510}{Phys. Rev. D \textbf{98}, 094510 (2018)}.

\end{thebibliography}
\end{document}